\documentclass[aps,prb,twocolumn,superscriptaddress,a4paper]{revtex4-2}
\usepackage[dvipdfmx]{graphicx}

\usepackage{color}
\usepackage{amsmath}
\usepackage{amssymb}
\usepackage{lipsum}
\begin{document}

\title{Mott transition and magnetism in a fragile topological insulator}

\author{Ashish Joshi}
\affiliation{Department of Physics, Kyoto University, Kyoto 606-8502, Japan}
\email[]{joshi.ashish.42a@st.kyoto-u.ac.jp}
\author{Robert Peters}
\affiliation{Department of Physics, Kyoto University, Kyoto 606-8502, Japan}

\begin{abstract}
We study the effects of electronic correlations on fragile topology using dynamical mean-field theory. Fragile topological insulators (FTIs) offer obstruction to the formation of exponentially localized Wannier functions, but they can be trivialized by adding certain trivial degrees of freedom. For the same reason, FTIs do not host symmetry-protected flow of edge states between bulk bands in cylindrical boundary conditions but are expected to have a spectral flow between the fragile bands and other bands under certain twisted boundary conditions.
We here analyze commonly observed effects of strong correlations, such as the Mott-insulator transition and magnetism, on a known model hosting fragile topology. We show that in the nonmagnetic case, fragile topology, along with the twisted boundary states, is stable with interactions below a critical interaction strength. Above this interaction strength, a transition to the Mott insulating phase occurs, and the twisted boundary states disappear. Furthermore, by applying a homogeneous magnetic field, the fragile topology is destroyed. However, we show that a magnetic field can induce a topological phase transition which converts a fragile topological insulator to a Chern insulator. Finally, we study ferromagnetic solutions of the fragile topological model.
\end{abstract}

\maketitle

\section{Introduction}
Stable topological insulators (STIs) are defined by the fact that they cannot be continuously deformed into the atomic limit without closing the bulk gap or breaking the underlying symmetries \cite{brouder_exponential_2007,budich_search_2014,shiozaki_topological_2017,read_compactly-supported_2017,chen_impossibility_2014,hasan_colloquium_2010}. Recently, a new class of topological insulators, dubbed as 'fragile topological insulators' (FTIs), has been discovered whose nontrivial bands, like STIs, do not allow the formation of exponentially localized Wannier functions preserving all symmetries, i.e. they offer Wannier obstructions denoted as 'non-Wannierizable'. But in contrast to STIs, an FTI can be adiabatically changed into an atomic insulator by adding a certain trivial set of bands, like that of a trivial band insulator \cite{po_fragile_2018,bradlyn_disconnected_2019,bouhon_wilson_2019,PhysRevX.10.031001,song_real_2020,peri_experimental_2020,else_fragile_2019}. Furthermore, while stable topological bands can be characterized by a nontrivial topological invariant, like a Chern number or a $\mathbb{Z}_2$ index, fragile topological bands cannot as they have trivial values for these topological invariants. 

Another peculiar feature of fragile topological phases is that they do not, in general, host symmetry-protected gapless surface or edge states. Since FTIs can be trivialized by the addition of atomic insulators without any surface or edge states while maintaining the bulk gap, there is no bulk-edge correspondence. However, Song et al. showed that there exists a new type of 'twisted' bulk-boundary correspondence for FTIs \cite{song_real_2020}. When a set of fragile topological bands are taken through some specific twisted boundary conditions (TBCs), symmetry-protected spectral flow exists between the fragile bands and other bands as a function of a single parameter ($\lambda$), which controls the deformation of these TBCs. Thus, traditional methods of characterizing topological insulators, like the bulk-boundary correspondence and topological indices, would label FTIs as topologically trivial. Thus, in addition to these methods, one also has to check whether each set of isolated bands is Wannierizable, or equivalently, if the Wilson loops of those bands do not wind \cite{bradlyn_disconnected_2019,bouhon_wilson_2019,hwang_fragile_2019,alexandradinata_wilson-loop_2014,wieder_axion_2018}. If this is not the case, the set of bands host nontrivial topology. The twisted bulk boundary correspondence also provides a way to measure the effects of fragile topology theoretically as well as experimentally, as has been recently realized in an acoustic metamaterial \cite{peri_experimental_2020}.

However, fragile topology is not only a theoretical peculiarity but also has significant practical consequences. As an example, consider the tight-binding models of materials, which are obtained from the Wannier functions of relevant bands. If there is no nontrivial topological invariant, one may conclude that the set of bands is trivial and hence Wannierizable. However, the bands could also be fragile topological, which would result in an obstruction to the formation of Wannier functions. Thus, it is essential to diagnose the topology in the bands of interest properly. Otherwise, the construction of a tight-binding model obeying the symmetries of the material might fail. Furthermore, fragile topology is predicted to be present in the band structures of a large number of real materials, including in the flat-bands of twisted bilayer graphene \cite{po_faithful_2019,ahn_failure_2019,song_all_2019,zou_band_2018,PhysRevX.10.031001,peri_fragile_2021}. 
In particular, the authors in Ref. \cite{peri_fragile_2021} studied the effects of attractive Hubbard interactions in a flat band model with fragile topology, with properties similar to the flat bands in twisted bilayer graphene. They showed that fragile topology plays an important role in achieving superconductivity.

Fragile topology was first realized in a noninteracting model given by Po et al. \cite{po_fragile_2018}, and much work has been done in understanding these systems in the noninteracting regime. In interacting systems, however, there have been relatively fewer studies \cite{else_fragile_2019,liu_shift_2019,latimer_correlated_2021,peri_fragile_2021}. On the other hand, STIs have been extensively studied with interactions and are predicted to show unconventional correlated topological states \cite{tada_study_2012,rachel_interacting_2018,hohenadler_correlation_2013}. Fractional Chern insulators, topological Mott insulators, topological Kondo insulators, and topological insulators without edge states are some of the examples of novel phases produced by the interplay of nontrivial topology and strong interactions \cite{maciejko_fractionalized_2015,neupert_fractional_2015,bergholtz_topological_2013,dzero_topological_2016,alexandrov_cubic_2013,wu_puzzle_2017,raghu_topological_2008,pesin_mott_2010,gurarie_single_2011,PhysRevB.98.075104}. In the case of FTIs, Ref. \cite{else_fragile_2019} discusses that certain interacting fragile topological phases are stable as long as the spatial symmetries are maintained and no additional set of bands is introduced in the system.
Interestingly, there are examples of FTIs that cannot exist in noninteracting regime but only in interacting systems \cite{else_fragile_2019,latimer_correlated_2021}. These FTIs require strong electron correlations as a necessary ingredient for many-body entanglement in their ground states.
The already mentioned study in Ref. \cite{peri_fragile_2021} considers attractive interactions in a flat band model to investigate its superconducting properties.

The effects of repulsive electron-electron correlations in FTIs have not yet been explicitly studied and it is still an open question whether electron correlations in FTIs can lead to nontrivial physical properties or not. The effects of magnetism and magnetic fields on fragile topology have not been studied as well. It is also not clear up to what degree the new twisted bulk boundary correspondence in FTIs holds in the presence of interactions and/or magnetic fields. Since FTIs, in general, do not possess 'normal' edge states which connect the bulk bands (as seen in STIs), it would be of interest to study the evolution of edge states in the interacting/magnetic regime. In this work, we study the effects of electronic correlations in the time-reversal symmetric (TRS) FTI introduced in Ref. \cite{po_fragile_2018}. For this purpose, we use dynamical mean-field theory (DMFT) \cite{georges_dynamical_1996} with numerical renormalization group (NRG) \cite{wilson_renormalization_1975,bulla_numerical_2008,peters_numerical_2006} to solve the impurity model. We find that in the nonmagnetic case, the fragile topological phase stays stable until a critical interaction strength, after which a transition to topologically trivial Mott insulating phase occurs. We verify the presence of spectral flow under some specific TBCs and show that this bandgap crossing is stable under interactions until the transition to the Mott phase, after which these states disappear. On applying a constant magnetic field, we show that the fragile topological character is lost, and topological phase transitions (as a function of field strength) to a stable topology occur in different sets of bands. To analyze these topological phases, we use Wilson loops. Finally, we investigate how stable magnetic phases due to interactions affect the topology of the system, and we find a stable ferromagnetic ordering made up of two sets of Chern insulators as conduction and valence bands above a critical interaction strength. We argue that while we study the interaction effects on a particular model, most of our results should apply to general fragile topological phases as well.

The rest of this article is organized as follows: In Sec. II, we briefly describe the honeycomb lattice FTI model and the DMFT/NRG technique to study the effects of interactions. In Sec. III, we discuss the FTI to Mott insulator transition. In Sec. IV, we define the TBCs for the honeycomb lattice under TRS and show the evolution of the twisted boundary states under interactions. In Sec. V, we study the topological phase transitions due to an applied magnetic field using Wilson loops, and in Sec. VI, we analyze magnetic solutions of the DMFT/NRG calculations. Finally, we conclude in Sec. VII.

\section{Model and Method}
To study interaction effects on fragile topology, we use a four-band honeycomb lattice model first proposed by Po et al. in Ref. \cite{po_fragile_2018}. The full description of the model is given there, and here we only briefly describe it. The model consists of a spinful $p_z$ orbital centered on each site ($2b$ Wyckoff position)  of a  honeycomb lattice with the origin at the center of the hexagon (space group $p6mm$ with time-reversal symmetry). Thus, there are two atoms in one unit cell. Then, starting with the Kane-Mele model \cite{kane_z_2_2005}, the authors introduce elaborate spin-orbit couplings and long-range hoppings (up to fifth nearest-neighbor). This causes a band inversion at $\Gamma$ and renders the $\mathbb{Z}_2$ quantum spin hall index trivial. Additionally, the inversion symmetry is removed to give a zero Chern number and leaves the model without stable topology.

\begin{figure}
	\includegraphics[width=\linewidth]{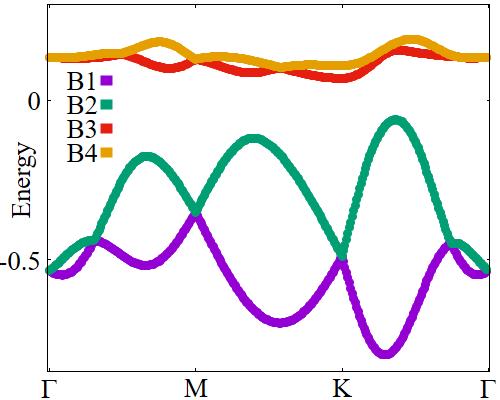}
	\caption{Band structure of the noninteracting FTI. The four bands are labelled $B1,B2,B3,B4$ from low energies to high energies. The valence bands ($B1$ and $B2$) are trivial while the conduction bands ($B3$ and $B4$) are fragile topological.}
	\label{Fig1}
\end{figure}

Explicitly, the model is constructed in the following way. Each time-reversal symmetric bond $i$ ($i=1,2,...,5$) (see Fig. 1(c) in Ref. \cite{po_fragile_2018}) is defined by a spin-independent hopping term $\tau_i$ and a spin-orbit interaction term $\boldsymbol{\gamma}_i$, and is given by
\begin{equation}
	\hat{h}_i\equiv \displaystyle\sum_{\alpha,\beta}\hat{c}_{\textbf{2},\alpha}^{\dagger}\left( \tau_i\sigma_0+i\boldsymbol{\gamma}_i\cdot\boldsymbol{\sigma}\right)_{\alpha\beta}\hat{c}_{\textbf{1},\beta}
\end{equation}
where $\hat{c}_{\textbf{1},\alpha}$ denotes the fermion annihilation operator acting on an electron with spin $\alpha$ at site \textbf{1}. $\sigma_j$ corresponds to the usual Pauli matrices. The values of the hopping term and the spin-orbit interaction term are real (necessary for the bonds to be TRS) and are given in Ref \cite{po_fragile_2018}. Now, summing over all the $g$ related bonds for $g\in G$ (where $G$ is the space group $p6mm$), i.e. summing over all the symmetrically equivalent bonds to a given bond $i$, gives the noninteracting Hamiltonian $\hat{H}_0$ as
\begin{equation}
	\hat{H}_0=\frac{t}{12}\displaystyle\sum_{i=1}^{5}\displaystyle\sum_{g\in G}\hat{g}\hat{h}_i\hat{g}^{-1} + \text{h.c.}
	\label{eq2}
\end{equation}
where we have used $t=1/3$ in this work. We have also shifted the band gap to $\omega=0$ by adding a constant. This is a four-band model, in which the lowest two bands (valence bands, $B1$ and $B2$) allow the formation of symmetric localized Wannier functions, whereas the other two bands (conduction bands, $B3$ and $B4$) are topologically nontrivial and offer obstruction to the formation of localized Wannier functions (Fig. \ref{Fig1}). Wannier obstructions can also be inferred from the Wilson loop windings \cite{alexandradinata_wilson-loop_2014,bouhon_wilson_2019,bradlyn_disconnected_2019,hwang_fragile_2019,wieder_axion_2018}. Thus, with a set of trivial valence bands but nontrivial conduction bands, this model shows no stable topology but fragile topology. It can be shown that on the addition of a certain trivial set of bands, the Wannier obstruction of the conduction bands can be trivialized, and the FTI can be converted to the atomic limit adiabatically \cite{else_fragile_2019}.
We study the interacting Hamiltonian $\hat{H}=\hat{H}_0 + \hat{H}_U$, where $\hat{H}_0$ is the noninteracting part defined in Eq. \ref{eq2} and $\hat{H}_U$ is the Hubbard interaction term given by 
\begin{equation}
	\hat{H}_U=	U\displaystyle\sum_j\hat{n}_{j\uparrow}\hat{n}_{j\downarrow}
\end{equation}
where $U$ is the on-site interaction strength and $\hat{n}_{j\alpha}$ is the particle number operator for an electron with spin $\alpha$ located at site $j$.

We use DMFT with NRG to investigate the physical properties of the full Hamiltonian $\hat{H}$. DMFT is a non-perturbative technique to study electron correlations \cite{georges_dynamical_1996} and has been successfully used in the context of STIs \cite{tada_study_2012,vanhala_topological_2016,kumar_interaction-induced_2016,irsigler_spin-imbalance-induced_2019,rachel_topological_2010,yoshida_correlation_2012,yoshida_topological_2013}. It includes local quantum fluctuations exactly and maps the many-body lattice problem to a quantum impurity model. Due to the mapping on an impurity model, DMFT can only describe effects of a momentum-independent self-energy, such as a renormalization of the band structure, Mott insulating behavior, energy shifts of bands, and magnetism. However, DMFT cannot describe long-range correlations or long-range entanglement. The impurity model is then solved self-consistently, starting with an initial guess for the self-energy. For this purpose, we use the NRG method, which was specially designed to solve quantum impurity models \cite{wilson_renormalization_1975,bulla_numerical_2008,peters_numerical_2006}. NRG can calculate highly resolved Green's functions near the Fermi energy but the resolution decreases away from the Fermi energy. However, this does not affect our results as long as the gap does not close. In the next section, we discuss the results of the DMFT/NRG calculations.

\section{Mott Transition}
\begin{figure}
	\includegraphics[width=\linewidth]{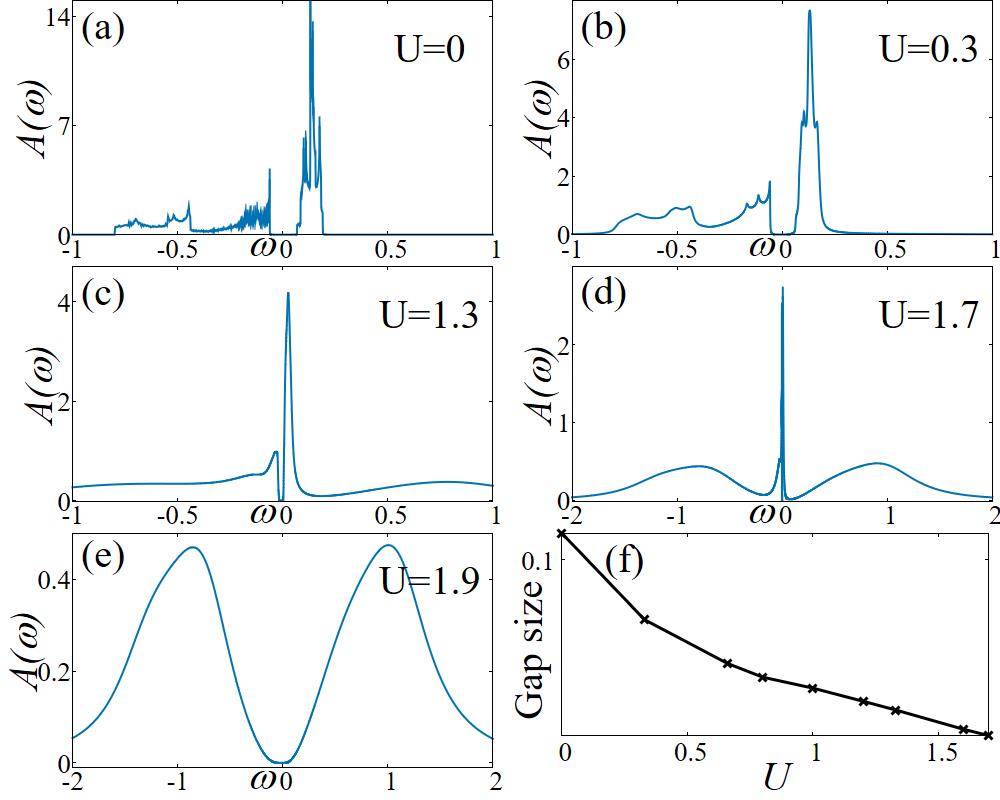}
	\caption{(a)-(e) Density of states, $A(\omega)$, as a function of frequency $\omega$ for different $U$. (f) Evolution of the band-gap with $U$ until the Mott insulator transition.}
	\label{Fig2}
\end{figure}

First, we discuss nonmagnetic solutions of the DMFT/NRG calculations performed on a homogeneous infinite lattice.
We calculate the density of states (DOS) of the $i$th band, $A_{ii}(\omega)$, for our model using
\begin{equation}
	A_{ii}(\omega)= -\frac{1}{\pi}\text{Im}\int d\mathbf{k}G_{ii}(\omega,\mathbf{k})
	\label{eq4}
\end{equation}
where the integration is performed over the full Brillouin zone (BZ) and $G(\omega,\mathbf{k})$ is the matrix of the single-particle Green's function, which is given by
\begin{equation}
	G_{ij}(\omega,\mathbf{k})=(\omega+i\delta+\mu-H_0(\mathbf{k})-\Sigma(\omega))^{-1}_{ij}
\end{equation}
where $\mu$ is the chemical potential and $\Sigma(\omega)$ is the $\omega$-dependent self-energy. The off-diagonal elements of the local Green's function between both lattice sites vanish, and we find that the diagonal elements are equal. Thus, the DOS and self-energy of both sites for spin-up and spin-down electrons are the same and we show the DOS of only a single band in Fig. \ref{Fig2}. Fig. \ref{Fig2}(a) shows the noninteracting DOS with a gap at the Fermi energy. 

\begin{figure*}
	\centering
	\includegraphics[width=\textwidth]{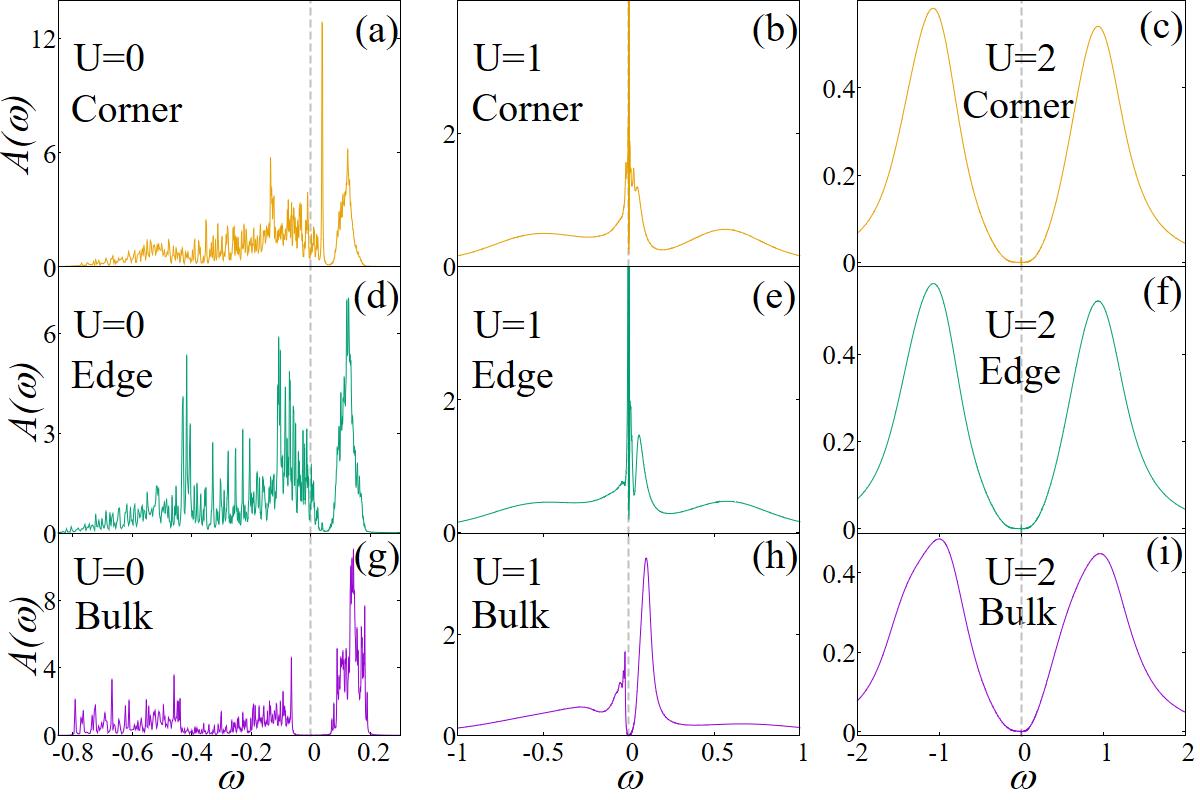}
	\caption{Edge and bulk local density of states at different interaction strengths. (a)-(c) Local DOS for the corner lattice site marked yellow in Fig. \ref{Fig4}(a). (d)-(f) Local DOS for the edge lattice site marked green in Fig. \ref{Fig4}(a). (g)-(i) Local DOS for the bulk lattice site marked purple in Fig. \ref{Fig4}(a).}
	\label{Fig3}
\end{figure*}

Now, we turn on the interaction while keeping $\mu=U/2$ for the system to be half-filled at all times (Fig. \ref{Fig2}(b)-\ref{Fig2}(e)). For small $U$ ($U=0.3$), the self-energy is small, and the DOS is similar to the noninteracting case.  On increasing $U$ further, we see the effects of the interactions as the DOS gets renormalized more strongly with $U$ and Hubbard bands emerge. Until a critical interaction strength ($U=U_c$), the bandgap decreases monotonically with increasing $U$ (Fig. \ref{Fig2}(f)). At $U=U_c$, the phase transition to a topologically trivial Mott insulating phase takes place. That this is indeed a Mott transition can be verified by the appearance of poles near the Fermi energy in the imaginary part of $\Sigma(\omega)$. On increasing $U$ further, the Mott bulk gap increases monotonously. A coexistence region exists between $U=1.6$ and $U=2.0$ in which both the fragile phase and the Mott phase are present, indicating that this is a first-order phase transition. The critical interaction strength $U_c$ lies in this coexistence region. Since we did not add any additional trivial degrees of freedom, neither broke any symmetries of the FTI while also maintaining the bulk gap for $U<U_c$, we expect that the fragile topological nature of our model is stable against interactions till $U_c$. We verify this in the next section by studying the evolution of the boundary states with interactions under TBCs.

\begin{figure}
	\centering
	\begin{minipage}[b]{0.45\linewidth}
		\includegraphics[width=\linewidth]{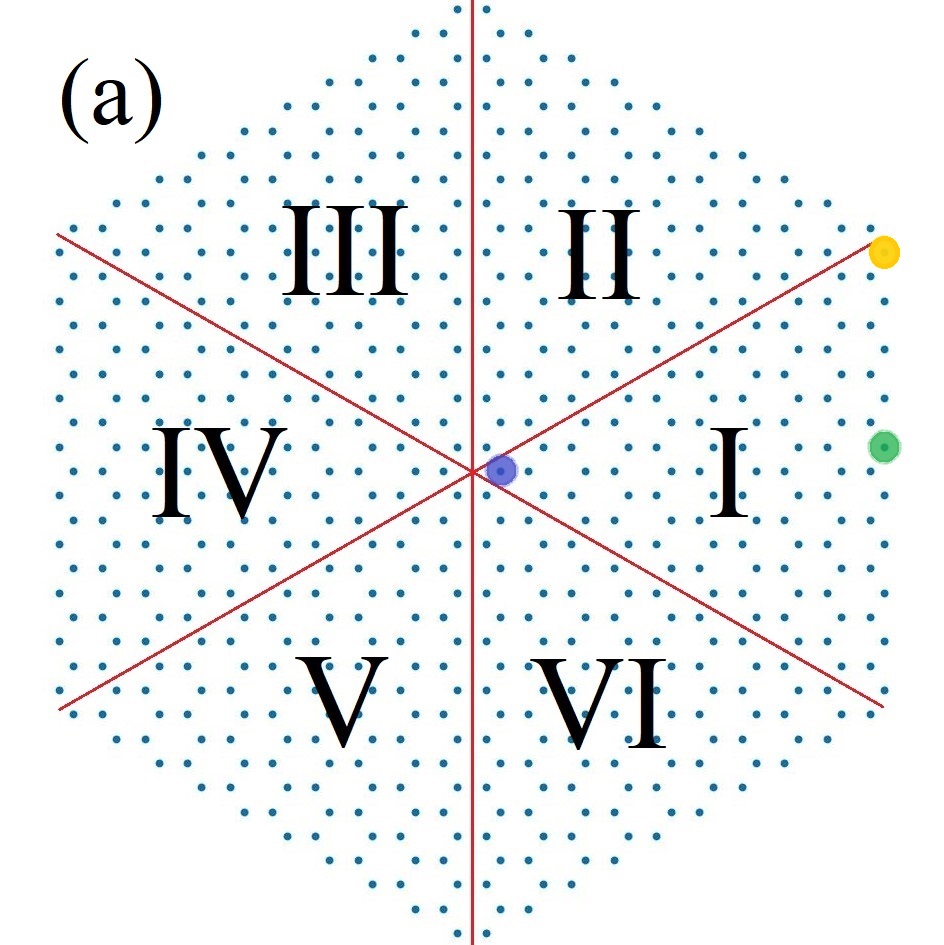}
	\end{minipage}	
	\begin{minipage}[b]{0.53\linewidth}
		\includegraphics[width=\linewidth]{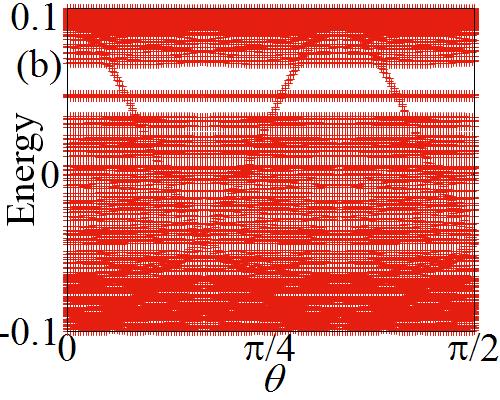}
	\end{minipage}
	\begin{minipage}[b]{0.49\linewidth}
		\includegraphics[width=\linewidth]{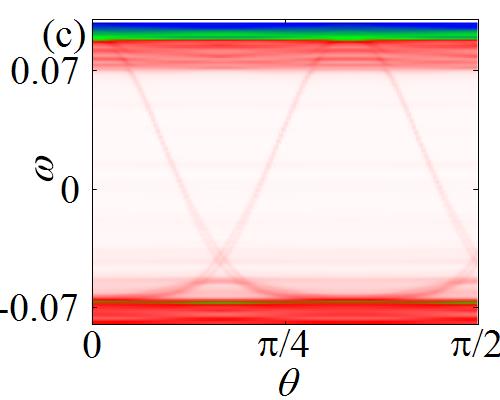}
	\end{minipage}	
	\begin{minipage}[b]{0.49\linewidth}
		\includegraphics[width=\linewidth]{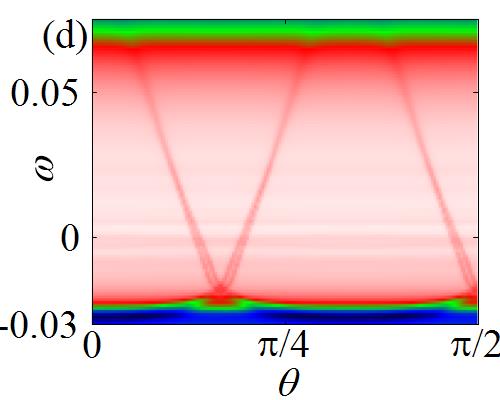}
	\end{minipage}	
	\begin{minipage}[b]{0.49\linewidth}
		\includegraphics[width=\linewidth]{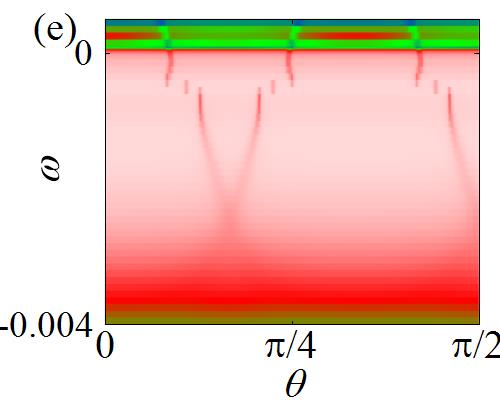}
	\end{minipage}
	\begin{minipage}[b]{0.49\linewidth}
		\includegraphics[width=\linewidth]{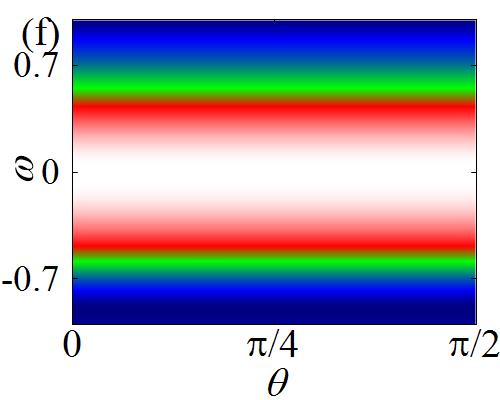}
	\end{minipage}
\caption{(a) A schematic diagram of the hexagonal lattice used in real-space calculations. Blue dots are lattice sites, and red lines divide the lattice into six $C_6$ equivalent sections. Hoppings that cross a red line from one section to another is multiplied by a phase factor according to the TBCs described in the text. (b) Spectral flow under TBCs in the noninteracting case. Unchanging states with $\theta$ are the localized in-gap edge states. (c)-(f) Spectral flow under TBCs for $U=0,0.6,1.7,2.0$ after removing the contribution of edge states for clarity.}
\label{Fig4}
\end{figure}

To study the edge states and evolution of the twisted boundary states with interactions, we translate our model onto a finite hexagonal flake with 1350 lattice sites. Fig.~\ref{Fig3} shows the local Green's functions at the bulk and the edges and Fig.~\ref{Fig4}(a)  shows the schematic diagram of the hexagonal flake used in the calculations. First, we discuss the noninteracting local Green's function for the bulk (at the center) and edge sites with open boundaries. While the bulk DOS (Fig. \ref{Fig3}(g))looks similar to the DOS of the homogeneous infinite lattice, the DOS at the edges shows a smaller gap due to the presence of many in-gap edge states (Figs. \ref{Fig3}(a) and \ref{Fig3}(d)). The reason for the occurrence of these in-gap states on the edges can be attributed to the 'filling anomaly' of fragile topology \cite{benalcazar_quantization_2019,song_real_2020}. Filling anomaly is the emergence of partially filled states at the edges or corner of a topological crystalline insulator due to a difference in the number of electrons present in an energy band and the number of electrons required for charge neutrality. Since an FTI does not display a flow of edge states traversing the bulk gap, these in-gap states are localized on the edges.

We now perform real-space DMFT/NRG calculations on our FTI model translated on a finite lattice with open boundary conditions \cite{titvinidze_dynamical_2012,irsigler_interacting_2019,snoek_antiferromagnetic_2008,PhysRevB.89.155134,PhysRevB.92.075103,wu_quantum_2012}. In real-space DMFT, all the geometrically inequivalent sites are mapped to their corresponding quantum impurity models, which we then solve self-consistently using NRG. We calculate the local DOS using Eq.~\ref{eq4} at each lattice site. Since the self-energy is now site-dependent, the edge sites experience the effects of interactions more strongly than the bulk (sites around the center of the lattice) \cite{tada_study_2012,PhysRevLett.114.177202,PhysRevB.93.235159}. The reason for the stronger correlation effects at the boundaries is the reduced coordination number at the edges compared to the bulk. The strength of correlation effects depends on the ratio between the interaction strength, $U$, and the kinetic energy given by the electron hoppings. While $U$ is local and does not change at the boundary, possible hopping processes are reduced at the boundaries compared to the bulk. The effect of this becomes especially prominent as we approach the Mott transition point. There, the interactions are large, and the gap in the edge states becomes unobservable due to stronger renormalization while the gap in the bulk remains until the critical interaction value, $U_c$, is reached (Figs. \ref{Fig3}(b), \ref{Fig3}(e) and \ref{Fig3}(h)). Nevertheless, this does not change the fragile topological nature of this system, and, under cylindrical boundary conditions, we do not find any bandgap crossings due to edge states present at the hexagonal edges. As we go beyond $U_c$, both the edge and bulk states show a simultaneous phase transition to a topologically trivial Mott insulating phase which is homogenous over the whole lattice (Figs. \ref{Fig3}(c), \ref{Fig3}(f) and \ref{Fig3}(i)).

\section{Twisted Boundary Conditions}
Now, we use TBCs for this hexagonal lattice. For a system defined on a torus under TBCs, when a particle crosses the twisted boundary, it gains a phase of $re^{i\theta}$ with $0\leq r\leq1$ \cite{qi_general_2006,song_real_2020}. We implement the TBCs by first dividing the hexagonal lattice flake into six equal sections, equivalent to each other under $C_6$. Then, the hoppings between a site from one section to a site in another section are modified by a factor $\lambda$ ($=e^{i\theta}$) while the hoppings that take place within the same section are unaltered (the system still has the usual open boundaries at the hexagonal edges). Thus, under TBCs, we obtain a $\theta$-resolved spectrum similar to a momentum resolved spectrum in the case of cylindrical geometry. Physically, the TBCs are gauge equivalent to a flux threading the torus, with $\theta$ being the phase due to flux threading \cite{qi_general_2006}. Ref. \cite{song_real_2020} gives the full details of all TBCs in 2D for all wallpaper groups. Here we only give the $C_6$ symmetric TBCs we use in our model. We modify the original Hamiltonian $\hat{H}(1)$ in section $x$ to a twisted Hamiltonian $\hat{H}(\lambda)$  in section $y$ as:
\begin{align*}
		\langle x,i|\hat{H}(\lambda)|y,j \rangle = &\\ 
		&\langle x,i|\hat{H}(1)|y,j \rangle, \:\:\:\:\:\:\:\:\:\:\:\:\:\:\:\:\:\: y=x \\
		&\lambda\langle x,i|\hat{H}(1)|y,j \rangle,\:\:\:\:\:\:\:\:\:\:\:\:\:\:\: y=x+1\\
		&\lambda^*\langle x,i|\hat{H}(1)|y,j \rangle, \:\:\:\:\:\:\:\:\:\:\:\:\: y=x-1 \\
		&\lambda^2\langle x,i|\hat{H}(1)|y,j \rangle, \:\:\:\:\:\:\:\:\:\:\:\:\: y=x+2 \\
		&\lambda^{*2}\langle x,i|\hat{H}(1)|y,j \rangle, \:\:\:\:\:\:\:\:\:\:\:  y=x-2\\
		&Re(\lambda^3)\langle x,i|\hat{H}(1)|y,j \rangle, \:\:\:\: y=x+3
\end{align*}
where, $|x,i\rangle$ is the $i$th orbital in the $x$th section ($x,y = $ I, II, III, IV, V, VI; see Fig. \ref{Fig4}(a)). Thus, the original and twisted Hamiltonians are equivalent up to a gauge transformation. Now, as we slowly change $\lambda$, the system will go through a gauge transformation. 
This transformation does not commute with $C_6$ and will, in general, change the energy eigenvalues. Only at the specific values of $\theta = 2\pi/6$, $C_6$ is conserved. For these $\theta$ values, the energy eigenvalues of $\hat{H}(\lambda)$ are equal to those of $\hat{H}(1)$ but the $C_6$ eigenvalues, along with the real space invariants (RSIs), are different. This will result in a spectral flow with the states at the $C_6$-center closing the gap. 
RSIs are thereby local good quantum numbers defined in the real space and protected by point group symmetries. RSIs can be calculated by symmetry eigenvalues of the band structure (for more details, please refer to \cite{song_real_2020} and the supplementary material therein).

Fig. \ref{Fig4}(b) shows the spectral flow of the twisted boundary states in the noninteracting case as $\theta$ ($\lambda=e^{i\theta}$) is varied from $0$ to $\pi/2$ (the bulk gap lies at $-0.06\leq E\leq0.07$). While there is no sign of symmetry-protected edge states in our model under the usual cylindrical geometry (periodic boundaries in one direction and open boundaries in the other), the twisted boundary states traversing the bulk gap under TBCs exhibit a direct consequence of the nontrivial fragile topology. In addition to the twisted boundary states, we also see numerous localized in-gap states which are not affected by the TBCs. These states are the same in-gap states localized at the edges of our hexagonal flake and are a result of the filling anomaly. Finally, as mentioned earlier, we find that these twisted boundary states emanate from the center of the lattice and are protected by the $C_6$ symmetry.

To analyze the evolution of the twisted boundary states with interactions, we plot the DOS $A(\omega,\theta)=-\text{Tr}[\text{Im}G(w,\theta)]/\pi$ as a function of $\theta$ and $\omega$ (Figs. \ref{Fig4}(c)-\ref{Fig4}(f)). The $\theta$ dependence of the DOS comes from the transformation of the Hamiltonian through TBCs. We use the local on-site self-energies obtained from the real-space DMFT/NRG calculations in evaluating the Green's function $G(\omega,\theta)$. Since the edge states experience stronger renormalization than the bulk, as we discussed earlier, the self-energies of the bulk and edge sites generally differ. But as the twisted boundary states originate from the bulk of the system, only the bulk self-energies can effectively alter the spectral flow. As the bulk gap reduces with increasing $U$ (due to stronger renormalization), the gap crossing under TBCs persists. Even close to the critical interaction strength $U_c$ when the bulk gap is very small, we can still see the signatures of twisted boundary states (Fig. \ref{Fig4}(e)). Finally, the bulk gap changes to a Mott gap and the twisted boundary states disappear with it. After that, the bulk gap increases with increasing $U$, but no twisted boundary states reemerge, and a homogeneous Mott insulating phase is found (Fig. \ref{Fig4}(f)). Similar behavior but in the case of stable topology and its edge states has been observed in previous studies \cite{tada_study_2012,yu_mott_2011,hohenadler_correlation_2013}. We also find that this spectral flow, and thus the fragile topological phase, is stable with doping as well until the Mott transition (not shown here). 

We note that twisted boundary conditions can be realized in experiments. In meta-materials, for example, TBCs are controlled by tuning some mechanical parameters, as shown in a recent work by Peri et al. \cite{peri_experimental_2020}. We expect our results from DMFT/NRG calculations and spectral flow under TBCs to be quite general among FTIs in the presence of interactions.

\section{Transition to stable topology under magnetic field}

\begin{figure*}
	\centering
	\begin{minipage}[b]{0.32\linewidth}
		\includegraphics[width=\linewidth]{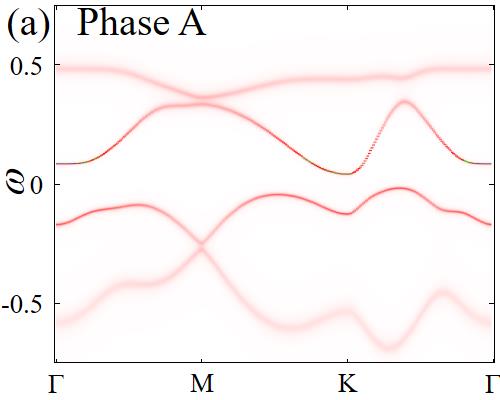}
	\end{minipage}	
	\begin{minipage}[b]{0.32\linewidth}
		\includegraphics[width=\linewidth]{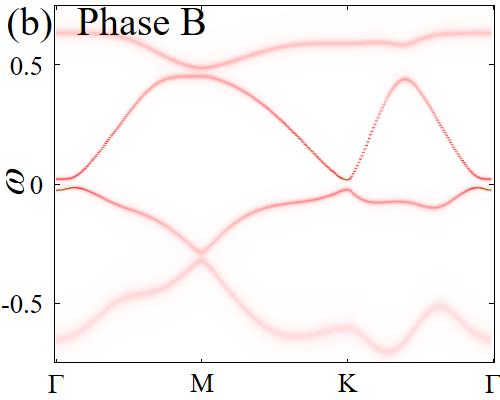}
	\end{minipage}
	\begin{minipage}[b]{0.32\linewidth}
		\includegraphics[width=\linewidth]{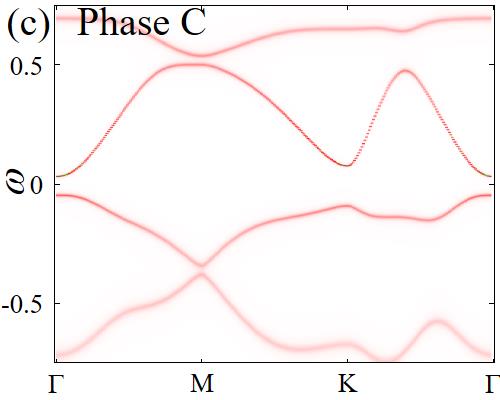}
	\end{minipage}	
	\begin{minipage}[b]{0.32\linewidth}
		\includegraphics[width=\linewidth]{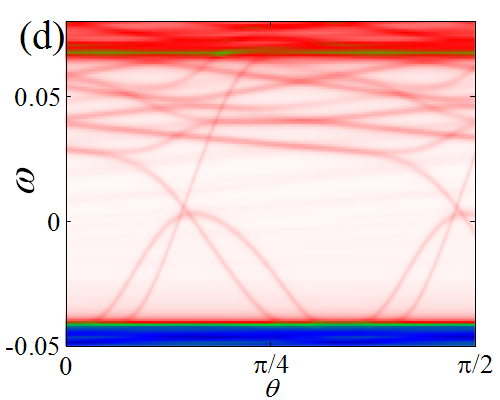}
	\end{minipage}	
	\begin{minipage}[b]{0.32\linewidth}
		\includegraphics[width=\linewidth]{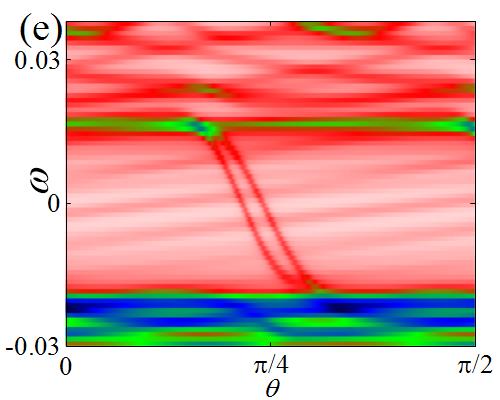}
	\end{minipage}
	\begin{minipage}[b]{0.32\linewidth}
		\includegraphics[width=\linewidth]{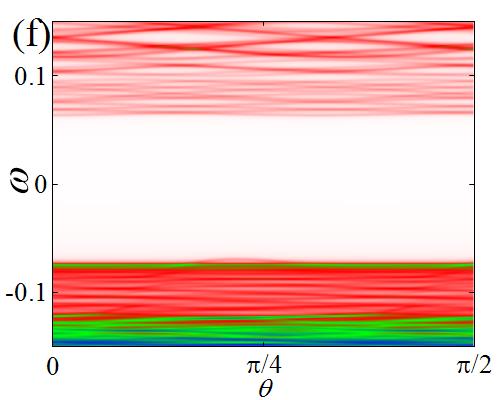}
	\end{minipage}
	\begin{minipage}[b]{0.32\linewidth}
		\includegraphics[width=\linewidth]{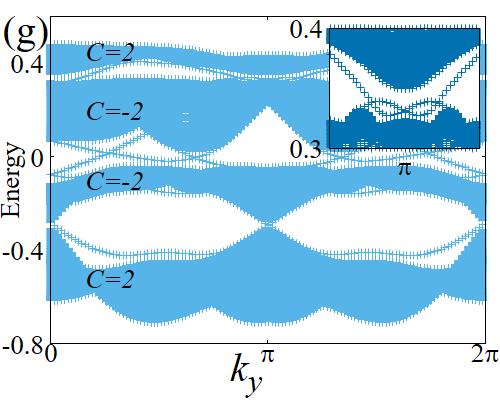}
	\end{minipage}	
	\begin{minipage}[b]{0.32\linewidth}
		\includegraphics[width=\linewidth]{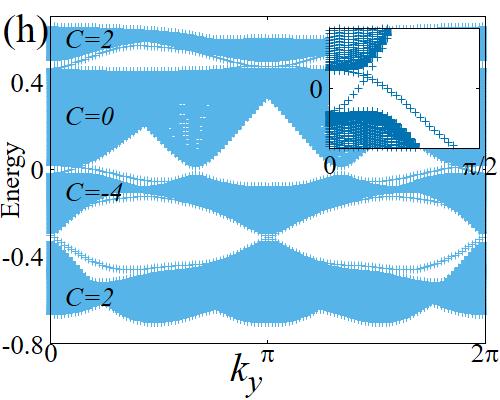}
	\end{minipage}
	\begin{minipage}[b]{0.32\linewidth}
		\includegraphics[width=\linewidth]{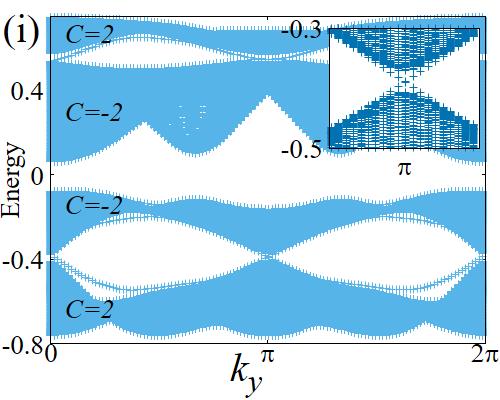}
	\end{minipage}	
	\caption{Left column: phase A, middle column: phase B, right column: phase C. (a)-(c) Band structures for the phases A, B, and C. (d)-(f) Phase A, B, and C under TBCs. (g)-(i) Phase A, B, and C diagonalized in a zig-zag strip geometry. $C$ gives the Chern number of a particular band. The insets show some of the edge states connecting bulk bands in the three gaps. All three phases have gap closing between $B1$ - $B2$ and $B3$ - $B4$. Phase B (h) also has edge states closing the gap between $B2$ and $B3$, which are lost in Phase C (i). All results are at $U=0.3$.}
	\label{Fig5}
\end{figure*}

In the last sections, we saw that fragile topology and twisted boundary states are robust under interactions until a critical point. Besides the Mott transition, magnetic ordering is a commonly observed phenomenon in strongly correlated materials. However, before studying a ferromagnetic state in this model of a fragile topological insulator, we want to analyze the effect of a constant homogeneous magnetic field applied perpendicularly to our FTI under interactions. In the case of STIs, presence of a magnetic field can result in exotic topological phenomena \cite{hasan_colloquium_2010,rachel_interacting_2018,tokura_magnetic_2019,morimoto_topological_2015,tran_impact_2020}. For example, a topological magneto-electric effect can be realized in a topological insulator in the event of a gap is created by a magnetic field \cite{hasan_colloquium_2010}. Here, we study the Hamiltonian $\hat{H}$ of our FTI in the presence of an external magnetic field of strength $b$, i.e., $\hat{H}-b\sigma_0\sigma_z/2$. Figs. \ref{Fig5}(a)-\ref{Fig5}(c) show the interacting band structure at different field strengths, $b$, for $U=0.3$. We identify the bands as $B1-B4$ from bottom to top. In the noninteracting and nonmagnetic case (Fig. \ref{Fig1}), the valence bands ($B1$ and $B2$) are connected at $\Gamma$, $M$ and $K$ while the conduction bands ($B3$ and $B4$) are connected at $\Gamma$ and $M$. For very small magnetic fields $b<b_0$, the band structure gaps for both sets of bands at $\Gamma$ and $K$ but stays connected at $M$. Increasing $b$ further opens up a gap at $M$ as well. Thus, for $b_0<b<b_1$, all the bands are disconnected at every $k$-point, and we call this region phase A.  Here $b_1$ is the critical field strength at which the bandgap between $B2$ and $B3$ closes again at $K$. On increasing $b$ further, the bandgap opens up again at all $k$-points and then closes for $b=b_2$ at $\Gamma$. We call the phase between $b_1<b<b_2$ phase B. For $b>b_2$, there is no more gap closing between any of the four bands, and the bandgap between $B2$ and $B3$ increases with $b$. We call this region phase C. Fig. \ref{Fig5}(a)-\ref{Fig5}(c) show the phases A, B, and C, respectively. The gaps at $M$ between $B1$ - $B2$ as well as $B3$ - $B4$ exist for all $b>b_0$. But due to the imaginary part of self-energy, it is smeared out in the interacting band structure and somewhat difficult to see in Figs. \ref{Fig5}(a)-\ref{Fig5}(c). For $U=0.3$ the values of $b_0, b_1$ and $b_2$ are $0.1,  0.36$ and $0.44$ respectively. These values of magnetic field strength for topological phase transitions decrease with increasing $U$.  

Figs. \ref{Fig5}(d)-\ref{Fig5}(f) show the three phases defined above under TBCs at $U=0.3$. In phase A (Fig. \ref{Fig5}(d)), two of the four twisted boundary states traversing the bulk gap in the noninteracting model gap out. There is still a spectral flow, but to find out whether the fragile topological nature of our model has survived, we use Wilson loops in the later part of this section. In phase B (Fig. \ref{Fig5}(e)), a new set of twisted boundary states emerge after closing and reopening of the bandgap. These twisted boundary states appear because of the emergence of stable topology between $B2$ and $B3$, as we will see later. Finally, in phase C (Fig. \ref{Fig5}(f)), the gap between $B2$ and $B3$ does not host any spectral flow as it has now been trivialized. Also, the fractional charges due to the filling anomaly have been neutralized, resulting in the disappearance of the in-gap localized edge states.

To characterize the topology in phases A,B, and C, we calculate the Chern number, $C_i$, for band $i$ using \cite{bernevig_topological_2013}
\begin{equation}
	\begin{split}
		&C_i=\frac{i}{2\pi}\int d\mathbf{k}\cdot\\
		&\displaystyle\sum_{j\neq i}\frac{\left\langle i(\mathbf{k})|\mathbf{\nabla_k}H'(\mathbf{k})|j(\mathbf{k})\right\rangle\times\left\langle j(\mathbf{k})|\mathbf{\nabla_k}H'(\mathbf{k})|i(\mathbf{k})\right\rangle}{(E_j(\mathbf{k})-E_i(\mathbf{k}))^2}
	\end{split}
	\label{eq6}
\end{equation}
where $|i(\mathbf{k})\rangle$ is the eigenstate of the $i$th band and $E_i$ is the $i$th eigenvalue. We use the effective topological Hamiltonian $H'(\mathbf{k})$ in Eq. \ref{eq6}, which is given by \cite{wang_simplified_2012}
\begin{equation}
	H'(\mathbf{k})=H_0(\mathbf{k})+\Sigma(\omega=0)
\end{equation}
and $\Sigma(\omega)$ is the momentum independent self-energy of the interacting Hamiltonian. This method only works when a smooth connection to the zero-frequency limit exists, which is the case for our model away from gap closing \cite{wang_simplified_2012}. Also, equation (6) is gauge independent as the differentiation is on the Hamiltonian and not on the wavefunction and does not explicitly depend on the phase of $|i\rangle$. 

We diagonalize our model in a zig-zag strip geometry with periodic boundary conditions in $y$-direction (along the momentum $k_y$) and open boundary conditions in $x$-direction (Fig. \ref{Fig5}(g)-\ref{Fig5}(i)) at $U=0.3$. In phase A (Fig. \ref{Fig5}(g)), gaps open up between the valence bands ($B1$ - $B2$) and conduction bands ($B3$ - $B4$). Each of these pairs of bands is now stably topological with Chern numbers $2,-2,-2,2$ for $B1,B2,B3,B4$ respectively. This also results in the emergence of multiple edge states crossing the bulk gap in each pair, associated with the emergence of stable topology in the bulk (inset of Fig. \ref{Fig5}(g) shows the edge states between $B3$ and $B4$ near $k_y=\pi$). The number of edge states between each pair of bands is equal to the difference in their corresponding Chern numbers. There is no stable topology between the bands $B2$ and $B3$ as they have the same Chern number. Though it may seem that there are edge states connecting these bands, these states do not form connections between both bands. In phase B (Fig. \ref{Fig5}(h)), there are edge states crossing the bulk gap between all bands, signifying the presence of stable topology with Chern numbers $2,-4,0,2$ corresponding to $B1,B2,B3,B4$ respectively. The inset of Fig. \ref{Fig5}(h) shows the crossings between $B2$ and $B3$ near $k=\pi/4$. Finally, in phase C, the middle gap is topologically trivial, but the other two gaps are nontrivial, with the same Chern numbers as in phase A (Fig. \ref{Fig5}(i)).

Thus, the new set of twisted boundary states in phase B is due to the stable topology between $B2$ and $B3$. Twisted boundary states also emerge in all three phases in the gaps between $B1$ - $B2$ and $B3$ - $B4$ as they now host stable topology. However, the gap between $B2$ - $B3$ in phase A does not host stable topology; the twisted boundary states in this phase indicate the presence of fragile topology. To confirm whether this is indeed a fragile topological phase, we diagnose the topology in these bands by calculating Wilson loops \cite{alexandradinata_wilson-loop_2014,bouhon_wilson_2019,bradlyn_disconnected_2019,hwang_fragile_2019,wieder_axion_2018},
\begin{equation}
	[W_{\mathbf{k_0+G\leftarrow k_0}}]^{ij}=\big\langle i(\mathbf{k_0 + G})\big| \displaystyle\prod_{\mathbf{k}}^{\mathbf{k_0+G\leftarrow k_0}}\mathcal{P}(\mathbf{k}) \big| j(\mathbf{k_0}) \big\rangle 
	\label{eq8}
\end{equation}
where $|i(\mathbf{k})\rangle$ is the eigenstate associated with the $i$th band and $\mathcal{P}(\mathbf{k})=\sum_{i}|i(\mathbf{k})\rangle\langle i(\mathbf{k})|$ is the projection operator to a certain set of bands. This set of bands must be energetically separated from other bands. The path-ordered product of $\mathcal{P}(\mathbf{k})$ is evaluated along the loop $\mathbf{k_0+G\leftarrow k_0}$ in the Brillouin zone. Windings of the Wilson loops can tell us about the topology of the selected set of bands- if it does not wind then the set of band is trivial. In the nonmagnetic case of our FTI model, the Wilson loop of conduction band winds but that of the valence bands does not wind \cite{bradlyn_disconnected_2019}. This implies that the conduction bands are non-Wannierizable, but the valence bands are Wannierizable. When we switch on the magnetic field, however, the Wilson loop windings are immediately lost in the conduction bands as well. Thus, the fragile topology in this model disappears in the presence of a magnetic field.

\begin{figure}
	\centering
	\begin{minipage}[b]{0.49\linewidth}
		\includegraphics[width=\linewidth]{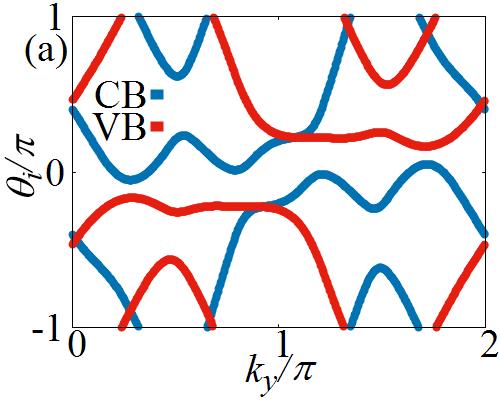}
	\end{minipage}	
	\begin{minipage}[b]{0.49\linewidth}
		\includegraphics[width=\linewidth]{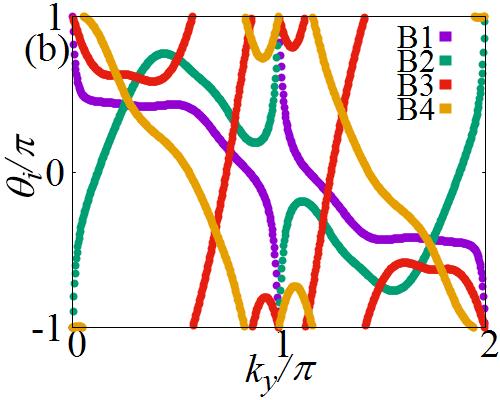}
	\end{minipage}
	\begin{minipage}[b]{0.49\linewidth}
		\includegraphics[width=\linewidth]{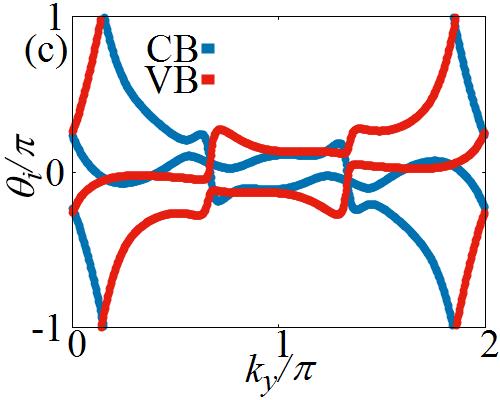}
	\end{minipage}	
	\begin{minipage}[b]{0.49\linewidth}
		\includegraphics[width=\linewidth]{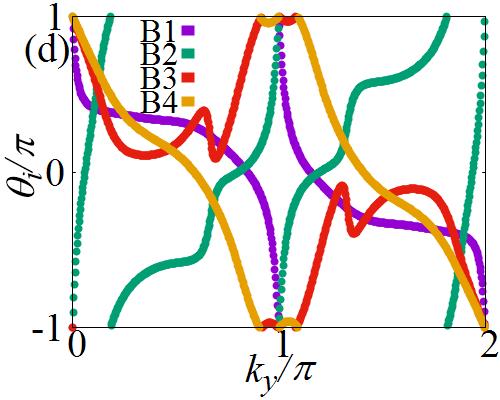}
	\end{minipage}	
	\begin{minipage}[b]{0.49\linewidth}
		\includegraphics[width=\linewidth]{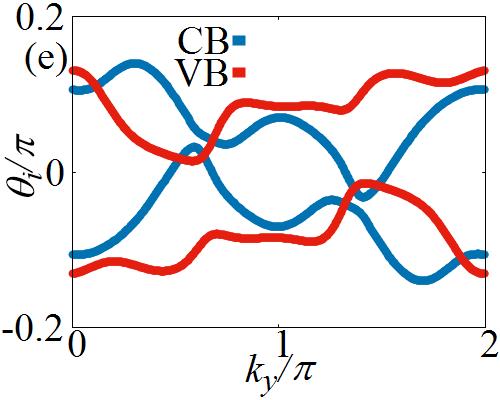}
	\end{minipage}
	\begin{minipage}[b]{0.49\linewidth}
		\includegraphics[width=\linewidth]{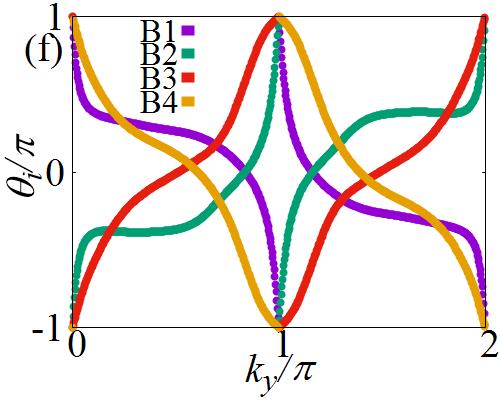}
	\end{minipage}
\caption{The figure shows the Wilson loops for the phases A, B, and C in the first, second, and third row, respectively. The left column shows the two-band Wilson loops, and the right column shows one-band Wilson loops. Both Wilson loops are calculated on a loop along $k_x$ at every $k_y$. $\theta_i$ is the argument of $i$-th eigenvalue of the Wilson loop matrix $W$. CB and VB denote Conduction Bands and Valence Bands, respectively, and $B1,B2,B3,B4$ are bands numbered from lowest to highest.}
\label{Fig6}
\end{figure}

We calculate the Wilson loops in two ways. First, by projecting $\mathcal{P}(\mathbf{k})$ on a set of two bands (valence bands or conduction bands), which we call the two-band Wilson loops. Second, by projecting $\mathcal{P}(\mathbf{k})$ on a single band (for all four bands), which we call the one-band Wilson loop. As calculating the Wilson loops by Eq. \ref{eq8} requires the set of bands to be isolated, one-band Wilson loop calculations are correct only when all the four bands are separated. This is indeed the case in our model for Phase A, B and C.
Fig. \ref{Fig6} shows the two-band and one-band Wilson loops for all three phases evaluated along $k_x$ at constant $k_y$, where $k_x$ and $k_y$ are the reciprocal lattice vectors of the BZ. This way, the whole BZ is traversed by $k_x$ and $k_y$ in computing $W$. We plot the argument of the $i$th eigenvalue of $W$, $\theta_i$, with respect to the momentum $k_y$. In phase A, the winding in Wilson loops (fig. \ref{Fig6}(a)) is lost in the presence of a magnetic field. This implies the absence of any nontrivial topology in the gap between the bands $B2$ and $B3$. The spectral flow under TBCs in phase A (Fig. \ref{Fig5}(d)) can be explained by the presence of non-zero RSIs which change on tuning the parameter $\theta$, indicating an obstructed atomic phase \cite{song_real_2020}. As expected, phase B shows a nontrivial winding in the two-band Wilson loop (Fig. \ref{Fig6}(c)) as the gap between the conduction and valence bands is stably topological. Phase C shows no winding (Fig. \ref{Fig6}(e)) of Wilson loops since there is no nontrivial topology in this case between the two sets of bands. Finally, we calculate the one-band Wilson loops in all three phases by projecting $\mathcal{P}(\mathbf{k})$ on only one band at a time (Figs. \ref{Fig6}(b), \ref{Fig6}(d) and \ref{Fig6}(f)). In all three phases we have individual bands well separated from each other away from the transition points. The windings in the one-band Wilson loop is due to the presence of nontrivial stable topology in the individual bands. The number of windings in the Wilson loops are equal to the Chern numbers of the respective bands. The sign of the Chern number can be deduced from the direction in which the Wilson loop winds. Thus, we corroborate the presence of nontrivial stable topology in all three phases and a loss of fragile topology in phase A using Wilson loops.

\section{Ferromagnetic Phase at high $U$}

\begin{figure}
	\centering
	\begin{minipage}[b]{0.49\linewidth}
		\includegraphics[width=\linewidth]{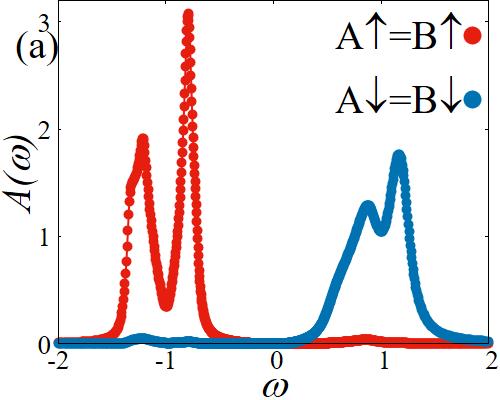}
	\end{minipage}	
	\begin{minipage}[b]{0.49\linewidth}
		\includegraphics[width=\linewidth]{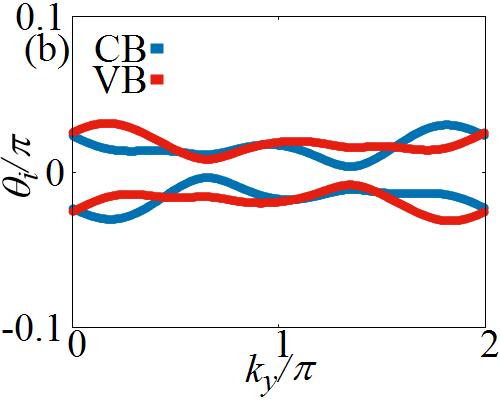}
	\end{minipage}
\caption{(a) Density of states of the FM phase at $U=2$. A and B denote the two sites of the hexagonal lattice; $\uparrow$ and $\downarrow$ correspond to spin up and spin down respectively. (b) Two-band Wilson loops for the same phase.}
\label{Fig7}
\end{figure}

A natural question now is whether our interacting FTI model can realize a stable magnetic phase in the absence of a magnetic field. If it can, it would be interesting to see whether the magnetic phase is topologically trivial or nontrivial. Thus, in this section, we study the magnetic solutions of our DMFT/NRG calculations. We first generate a small magnetic instability in our model and then study the possibility of a stable magnetic phase at zero temperature in the absence of any magnetic field. In section III, the self-energies of both lattice sites in the unit cell and both spin-directions were the same. Thus, we performed the DMFT/NRG calculations for only one impurity model per unit cell. Now, however, we solve two impurity models per unit cell corresponding to the two sites. Solving the DMFT equations for a homogeneous infinite lattice, we find that for any interaction strength less than a critical value ($U<U_{mc}$), the self-energy oscillates with every iteration; a self-consistent solution cannot be found. This indicates the possibility of a spin density wave (SDW) solution. For $U>U_{mc}$ however, we find a stable ferromagnetic (FM) phase (Fig. \ref{Fig7}(a)) which is strongly polarized (evident from the occupation numbers of spin-up ($n_\uparrow \approx 1$) and spin down ($n_\downarrow \approx 0$) electrons at each site (we start our calculations at the half-filling condition, $\mu=U/2 \implies n_\uparrow=n_\downarrow=0.5$)). In this strongly polarized FM phase, we do not find a spectral flow under TBCs. Also, the two-band Wilson loop does not wind (Fig. \ref{Fig7}(b)) while the one-band Wilson loops show the same windings as that of phase C obtained in Section V (Fig. \ref{Fig6}(f)). This signifies the absence of any topological character between the conduction and valence bands but the existence of stable topology between $B1$ - $B2$ and $B3$ - $B4$. The stable FM phase is then made up of two sets of Chern insulators separated by a trivial bandgap.

In the rest of this section, we discuss our solutions for $U<U_{mc}$. Since there are indications of an SDW phase, a two-site Hamiltonian is not enough to analyze the long-range ordered phase. Thus, we again use our hexagonal lattice flake and perform real-space DMFT/NRG calculations\cite{PhysRevB.89.155134,PhysRevB.92.075103}. To get a converged self-consistent solution for this potential SDW phase, we start by destabilizing only a few sites in the bulk of the lattice. We also average over self-energies of two successive iterations to prevent sudden spin flips, which may lead to an oscillating solution. We find that for $U<U_{mc1}$, where $U_{mc1}\approx1.3$, a homogenous paramagnetic phase exists at the bulk as well as at the edge sites. This paramagnetic phase retains the fragile topological nature of the original model as neither any symmetry is broken nor the bulk gap is closed. 
At $U=U_{mc1}$, while most of the lattice sites remain in the paramagnetic phase, the edge sites show random spin polarizations. This spin polarization can be attributed to the localized states emerging at the edges of the flake. A self-consistent solution cannot be found because these localized states make the system very sensitive to numerical errors. This 'flipping' of spins in one or the other direction begins at the edges and makes its way towards the center of the lattice as $U$ is increased further. However, at $U\geq U_{mc2}$, where $U_{mc2}\approx2$, most of the lattice sites are aligned in one direction, yielding an FM phase obtained earlier in the calculations for the homogeneous infinite lattice. Thus, we conclude that there is no magnetic solution for weak interactions, but a topologically nontrivial FM phase exists for $U>U_{mc}$ which is composed of two Chern insulators separated by a trivial gap at Fermi energy.

\section{Summary}

In this work, we have explicitly studied the effects of electron correlations and external magnetic fields on the recently discovered fragile topology and the twisted boundary states associated with it using DMFT/NRG. We have found that interactions do not destroy fragile topology until a critical interaction strength that triggers the Mott transition is reached. Twisted boundary states also show the same behavior and are not present in the Mott insulating phase. We have shown that the edges experience the effects of correlations more strongly than the bulk due to a lower coordination number. This results in a qualitatively different DOS at the edges and the bulk. But this effect has no bearing on the fragile nature of the model, and the Mott transition is also homogeneous over the whole lattice. We then switched on a constant external magnetic field which instantly destroys the fragile topology. We have shown that topological phase transitions can take place as a function of the magnetic field strength, converting a fragile topological phase to a stable Chern insulating phase. Diagnosis of topology is done using Wilson loops and Chern numbers. Finally, we have shown that a topologically nontrivial FM phase, composed of two sets of Chern insulating bands, is stable above a critical interaction strength in the absence of a magnetic field. While the noninteracting model is a fragile topological insulator, correlations, particularly magnetism, can change it to a stable topological insulator, which is an exciting prospect in future studies of correlated fragile insulators.

\textit{Note added:} We came across Ref. \cite{bouhon_topological_2020} while finalizing this work. They have also found a link between the magnetic fragile phase and the Chern insulator phase. We thank the authors for bringing their recent work to our attention.

\begin{acknowledgments}
We thank Y. Yanase for fruitful discussions. A. J. is supported by the MEXT Scholarship. R.P. is supported by JSPS, KAKENHI Grant No. JP18K03511. 
Parts of the computation in this work have been done using the facilities of the Supercomputer Center, the Institute for Solid State Physics, the University of Tokyo.
\end{acknowledgments}

\bibliography{references}

\end{document}